\documentstyle[12pt]{article}

\def\abstract#1{\vskip 7mm 
	\begin{center}{\large Abstract}\par \bigskip
		\begin{minipage}[c]{12cm}
			\small #1
		\end{minipage}
	\end{center}
}
\def\title#1{\begin{center}{\Large\bf #1}\end{center}}
\def\author#1{\vskip 5mm \begin{center}{#1}\end{center}}
\def\address#1{\begin{center}{\it #1}\end{center}}

\def\sech{{\rm sech}}

\begin{document}
\title{Magnetic solutions to 2+1 dimensional gravity with dilaton field
}
\author{Takao Koikawa\footnote{e-mail address: koikawa@otsuma.ac.jp}, \ Takuya Maki\footnote{e-mail address: maki@clas.kitasato-u.ac.jp}$^\dagger$  and Atsushi Nakamula\footnote{e-mail address: nakamula@sci.kitasato-u.ac.jp}$^\dagger$}
\vspace{1cm}
\address{
  School of Social Information Studies\\
         Otsuma Women's University\\
         Tama, Tokyo 206, Japan\\
\vspace{1cm}
$^\dagger$School of Science, Kitasato University\\ 
	Sagamihara, Kanagawa 228, Japan}
\vspace{2.5cm}
\abstract{
 We show a general method to solve 2+1 dimensional dilatonic Maxwell-Einstein equation with a positive or negative cosmological constant. All the physical solutions are listed with assumptions that they are static, rotationally symmetric, and has a nonzero magnetic field and a nonzero dilaton field. On the contrary to the magnetic solution without a dilaton field, some of the present solutions with a dilaton field possess a horizon.}
\newpage

The discovery of the black hole solution by Ba\~{n}ados, Teitelboim and Zanelli(BTZ) \cite{btz,bhtz,car} has generated a great interest in the 2+1 dimensional theory.
The solution was obtained assuming the negative cosmological constant and it can be formulated as the string theory\cite{howl,kal}. 
The  black  hole solution  would be a black string by the duality transformation.
Since the discovery by BTZ, many people have tried to extend the theory so that it could incorporate other fields such as the electromagnetic fields \cite{kako,cs} and the dilaton field \cite{cm,kp}. 
The characteristic feature of 2+1 dimensional theory with the electromagnetic field lies in its asymmetry of the electric fields and the magnetic field. 
While there are two electromagnetic field components, there is only one magnetic field in the 2+1 dimensional theory. 
This should be contrasted with the 3+1 dimensional theory where there exixt a symmetry between electric fields and magnetic fields.
Although there is no duality-like relation between the electric and magnetic fields, we can introduce an idea of duality in the subspace of the 2+1 dimensional spacetime \cite{kako}. 
Lately an exact solution with a magnetic charge is discussed in refs.\cite{kp,hw}.
In ref.\cite{hw}, the magnetic solution accompanied with negative cosmological constant was shown, and found that there is no horizon. 
In this sense this is a particle-like solution and might be called a monopole solution. 
It is well known that the introduction of the dilaton field drastically changes the spacetime structure in 3+1 dimensions. 
Therefore, also in 2+1 dimensions, it is worthwhile to pursue how the solutions might be modified when we incorporate a dilaton field. 
In ref.\cite{kp,pk} the solution was obtained by resorting to the duality transformation.
The solution including a magnetic charge was obtained from the solution with an electric charge discussed in \cite{cm}. 
Since the electric solution was constructed by adopting the ansatz on the dilaton field, the magnetic solution obtained by use of the duality transformation is just a class of exact solutions with a magnetic charge. 
We are not sure to which extent the solution is general. 
Therefore it is desirable to obtain solutions without assuming any ansatz except the rotational symmetry and static condition. 
The complete list of solutions would be useful to construct and understand the string theory connected by the duality.
These motivate the present work. 
We shall obtain the general magnetic solutions without resorting to the duality transformation in this paper. 
In order to get  general solutions we apply the method used to construct the solutions to the dilaton theories coupled to gravity and gauge fields \cite{ky}, where the solutions were exactly solved by introducing a new variable instead of the radial coordinate variable. The introduction of new variable simplifies the equations to a great extent. We apply the same technique to solve the present solution.

Consider a general metric class of massless dilaton theories coupled to gravity accompanied with a cosmological constant $\Lambda$ and $U(1)$ gauge field of which the action is given by
\begin{equation}
S=\int d^3x\sqrt{-g}\{R-{B\over2}(\partial\phi)^2-e^{4a\phi}F^2-2e^{b\phi}\Lambda\},
\end{equation}
where $a,\ b$ and $B$ are constants. 
We seek static solutions with rotational symmetry by introducing the coordinates
\begin{equation}
ds^2=-e^\sigma dt^2+e^\omega dr^2+r^2d\phi^2,
\end{equation}
where $\sigma$ and $\omega$ are functions of the radial coordinate variable $r$.
 In order to find a magnetically charged solution,  we assume that the only nonzero component of gauge potential is  $A_\phi$ which yields a magnetic field. 
The Einstein equation and the field equations lead to

\begin{eqnarray}
\sigma''+{1\over r}\sigma'+{1\over2}(\sigma'-\omega')\sigma'&=&4{e^{-4a\phi}\over r}(A'_\phi)^2-4e^{b\phi}e^\omega\Lambda,\label{3}\\
\sigma''+{1\over r}\omega'+{1\over2}(\sigma'-\omega')\sigma'&=&-B\phi'^2-4e^{b\phi}e^\omega\Lambda,\\
{1\over r}(\sigma'-\omega')&=&-4e^{b\phi}e^\omega\Lambda,\\
\phi''+{1\over r}\phi'+{1\over2}(\sigma'-\omega')\phi'&=&-{8a\over B}{e^{-4a\phi}\over r}(A'_\phi)^2+{2b\over B}e^{b\phi}e^\omega\Lambda,
\end{eqnarray}
\begin{equation}
A_\phi''+A_\phi'\{{1\over2}(\sigma'-\omega')-{1\over r}-4a\phi'\}=0,
\end{equation}
where the prime denotes an $r$-derivative.
We first integrate eq.(7) to obtain the magnetic field in the coordinate basis
\begin{equation}
{\cal B}=A_\phi'=qre^{4a\phi}e^{-(\sigma-\omega)/2},
\end{equation}
where $q$ is a constant of integration.
 
We have taken a unit such that $8\pi G=1$. 
This system of differential equation admits complete integration. 
We introduce a new variable $z$ by the following equation as in ref.\cite{ky}:
\begin{equation}
re^{(\sigma-\omega)/2}{dz\over dr}=1.
\end{equation}

The outline of solving the above system of differential equations is as follows.
 First we rewrite the above differential equations to those with $z$-derivatives, and then we integrate to obtain unknown functions in terms of the $z$ variable. 
Finally we connect $z$ with $r$ by integrating above definition of $z$.

The above equations can be rewritten by use of $z$ variables as
\begin{eqnarray}
\ddot\sigma&=&4r^2(q^2e^{4a\phi}-\Lambda e^{b\phi+\sigma}),\label{10}\\
\dot\sigma+\dot\omega&=&B\dot\phi^2e^{-(\sigma-\omega)/2}+4q^2r^2e^{4a\phi}e^{-(\sigma-\omega)/2},\label{11}\\
\dot\sigma-\dot\omega&=&-4\Lambda r^2e^{b\phi}e^{(\sigma+\omega)/2},\label{12}\\
\ddot\phi&=&-{2b\over B}r^2({4a\over b}q^2e^{4a\phi}-\Lambda e^{b\phi} ),
\end{eqnarray}
where the dotted fields represent derivatives with respect to $z$ and the two dotted the second order derivatives. 
Assuming $4a=b$, we obtain from eqs.(10) and (11)
\begin{equation}
\ddot\sigma+{2B\over b}\ddot\phi=0.
\end{equation}
Integrating this equation, we find that the linear combination of $\sigma$ and $\phi$ is expressed as 
\begin{equation}
\sigma+{2B\over b}\phi=c_1(z-z_1),
\end{equation}
where  $c_1$ and $z_1$ are constants of integration.
We shall next solve the equation for a linear combination of $\sigma$  and $\omega$.
 For this purpose we derive a second order differential equation for a function $f:=e^{(\sigma-\omega)/2}$.
Assuming $2B=b^2$, we get from (12) and  (15)

\begin{equation}
\dot{f}=(e^{(\sigma-\omega)/2})\dot{ }=-2\Lambda r^2 e^{c_1(z-z_1)}.\label{16}
\end{equation}
The use of definition of $z$ makes it possible to derive an equation only for $z$ 
variable:
\begin{equation}
\ddot f-2f\dot f-c_1\dot f=0.
\end{equation}
This is integrated to yield
\begin{equation}
\dot f=f^2+c_1f+k,
\end{equation}
where $k$ is a constant. 

In the integration of the above equation,  the solutions are classified in accordance with the value of $D:=c_1^2- 4k$ being (i)negative, (ii)zero or (iii)positive.  
In each case, we can easily integrate eq.(9) to express $r$ in terms of $z$.
 So far we have not specified the sign of the cosmological constant.
In  deriving the physical solutions in each case, we need to assign the sign.
We label the negative $\Lambda$ case as (I) and positive one as (II).
The solutions are classified by the values of $D$ and $\Lambda$.
For the  case(I-i), {\it i.e.} $\Lambda<0$ and $D<0$, we obtain
\begin{equation}
f=e^{(\sigma-\omega)/2}=-{c_1\over 2}-\sqrt{|c_2|\over 2}\cot\sqrt{|c_2|\over 2}(z-z_2),\label{19}
\end{equation}
where $c_2=2(c_1^2- 4k)<0$ and $z_2$ are constants. 
The radial coordinate $r$ is expressed in terms of $z$ as
\begin{equation}
r={1\over2}\sqrt{|c_2|\over |\Lambda|}{e^{-c_1(z-z_1)/2}\over\sin\sqrt{|c_2|\over 2}(z-z_2)}\label{20},
\end{equation}
where the range of $z$ is given by
\begin{equation}
\cot^{-1}(-{c_1\over\sqrt{2|c_2|}})\leq \sqrt{|c_2|\over2}(z-z_2)<\pi.
\end{equation} 
The range is obtained so that this guarantees the positive $f$.
For the  case (I-iii), {\it i.e.} $\Lambda<0$ and $D>0$,  the sinusoidal functions in above equations are to be taken place by hyperbolic function together with the replacement of negative $c_2$ by the positive one defined by $c_2=2(c_1^2- 4k_1)>0.$ The solutions in this case are to be obtained by simply replacing $c_2<0$ in eqs.(19) and (20) by $c_2>0$.
 For the  case (I-ii), {\it i.e.} $\Lambda<0$ and $D=0$,
\begin{equation}
f=e^{(\sigma-\omega)/2}={-{c_1\over 2}}-{1\over z-z_2}.
\end{equation}
The radial coordinate $r$ is given by 
\begin{equation}
r={e^{-c_1(z-z_1)/2}\over\sqrt{2|\Lambda|}(z-z_2)}.
\end{equation}
Here we also assume that the range of $z$ is limited from below by $z>z_2$.
The solutions here might be regarded as the  $c_2=0$ limit of the case (I-i).
 We can also obtain the solution for the case (II-iii), {\it i.e.} $\Lambda>0$ and $D>0$. 
The solution is given by
\begin{equation}
f=e^{(\sigma-\omega)/2}={-{c_1\over 2}}-\sqrt{c_2\over 2}\tanh\sqrt{c_2\over 2}(z-z_2),\label{23}
\end{equation}
where $c_2=2(c_1^2-4k)>0$. The radial coordinate $r$ is given by 
\begin{equation}
r={1\over2}\sqrt{|c_2|\over \Lambda}e^{-c_1(z-z_1)/2}\sech\sqrt{c_2\over 2}(z-z_2)
,
\end{equation}
where $z\geq z_2$ is assumed.
 There are no counterparts of (I-i) and (I-ii) in positive $\Lambda$ case because we can not have real $r$ in these cases.

We shall further solve the system of equations  for the case (I-i) and the method is also applicable to other cases.
Using $4a/b=1$, we obtain from eqs.(\ref{10}) and  (\ref{16})
\begin{equation}
\ddot{\sigma}=2(e^{(\sigma-\omega)/2})\dot{ }(1+{q^2\over|\Lambda|}e^{-\sigma})\label{25}.
\end{equation}
Substituting the solution (\ref{19}), we get the Liouville equation
\begin{equation}
\ddot{y}=e^{-y},\label{26}
\end{equation}
where $y$ is given by

\begin{equation}
y=\sigma+2\log|\sin\sqrt{|c_2|\over 2}(z-z_2)|-\log{|c_2|q^2\over|\Lambda|}.\label{27}
\end{equation}
We can easily integrate eq.(27) to get $\sigma$ as
\begin{equation}
\sigma=\log|{q^2|c_2|\over c_3|\Lambda|}{\cosh^2\sqrt{c_3\over 2}(z-z_3)\over\sin^2\sqrt{c_2\over 2}(z-z_2)}|,
\end{equation}
where $c_3$ and $z_3$ are constants of integration.
From this solution with (\ref{19}),  we obtain $\omega$ as
\begin{equation}
\omega=\log|{q^2|c_2|\over  c_3|\Lambda|}
{\cosh^2\sqrt{c_3\over 2}(z-z_3)\over\{-{c_1\over2}\sin\sqrt{c_2\over 2}(z-z_2)-{|c_2|\over2}\cos\sqrt{c_2\over 2}(z-z_2)\}^2}|.
\end{equation}
The use of (15) leads to expressing  the dilaton field by $z$ as
\begin{equation}
e^{b\phi}={c_3|\Lambda|\over q^2|c_2|}
{\sin^2\sqrt{{|c_2|\over2}}(z-z_2)\over\cosh^2\sqrt{c_3\over 2}(z-z_3)}e^{c_1(z-z_1)}.
\end{equation}
Finally  the magnetic field is also expressed in terms of $z$ as
\begin{eqnarray}
{\cal B}&=&-{c_3\over 2q}\sqrt{|\Lambda|\over |c_2|}{\sin^2\sqrt{|c_2|\over 2}(z-z_2)e^{{c_1\over2}(z-z_1)}\over \cosh^2\sqrt{c_3\over 2}(z-z_3)}\nonumber \\
& &\times\{ 
{c_1\over2}\sin\sqrt{|c_2|\over 2}(z-z_2)+\sqrt{|c_2|\over 2}\cos\sqrt{|c_2|\over 2}(z-z_2) \}^{-1}.
\end{eqnarray}

In the similar way we obtain the solutions for (I-ii), (I-iii) and (II-iii) cases. Of the Einstein equations, one remaining equation is  used to obtain a condition between constants appearing in solutions.
From eq.(\ref{11}) with eqs. (\ref{10}) and (\ref{12}), we get,
\begin{equation}
2\dot\sigma=(B\dot\phi^2+\ddot\sigma)e^{-(\sigma-\omega)/2},
\end{equation}
which brings about   
\begin{equation}
{c_1^2\over2}+c_3-|c_2|=0.
\end{equation}
In other cases, we get the same conditions with a slight modification of the sign of the constants.
For all cases (I) and (II)  $c_3>0$, and $c_2$ is  negative for (I-i), zero for (I-ii) and positive for (I-iii) and (II-iii).
 This shows that the case (I-ii), where $c_2=0$ in the above equation, should be excluded from our non-trivial solution lists.

We shall now discuss the allowed range of $r$ which comes from the positivity condition of $f.$ 
 In (I-i) case the allowed range of $r$ is given by
\begin{equation}
r_{\rm min}\leq r<\infty,
\end{equation}
corresponding to eq.(21).
Here $r_{\rm min}$ is given by
\begin{equation}
r_{\rm min}=\sqrt{|c_2|\over|\Lambda|}{e^{-c_1(z_2-z_1)/2}\over\sqrt{2|c_2|+c_1^2}}e^{{-c_1\over\sqrt{2|c_2|}}\cot^{-1}{(-c_1)\over\sqrt{2|c_2|}}}.
\end{equation}
We can easily show that $e^{-\omega}=0$ at $r_{\rm min}$, which indicates that the lower bound is the position where the horizon is located at.
In case (I-iii), we find that the allowed range of radial coordinate is all the positive region, and that $e^{-\omega}>0$ for the region.
This shows that the case does not fall into the solution with horizon, and we may call it a particle-like solution.
This may be interpreted as a monopole as in ref.\cite{hw} .
In the remaning case (II-iii), $r$ is  limited from above at the horizon.

 We shall now concentrate on (I-i) case because this is the only solution outside of the horizon, with asymptotic region.
Note that
\begin{equation}
r\longrightarrow\infty, \quad{\rm as} \quad \sqrt{|c_2|\over2}(z-z_2)\longrightarrow\pi.
\end{equation}
Then we find that the  asymptotic behavior of the metric is given by
\begin{eqnarray}
e^\sigma &\sim& r^2,\\
e^\omega&\sim& {\rm const.},
\end{eqnarray}
which does not coincide with that of BTZ solution with vanishing angular momentum, or anti-de Sitter spacetime.
In order to discuss the mass we follow the formalism developed by Brown and {\it et al.} \cite{by,bcm}.
The quasilocal mass is obtained as the conserved charge corresponding to the timelike Killing vector.
 In defining the quasilocal mass, there is an ambiguity coming from possible additional action at the boundary.
Here we assume no contribution from the boundary, and we get the quasilocal mass
\begin{equation}
M=-2e^{(\sigma-\omega)/2},
\end{equation}
 which is proportional to $f$. We then find that the quasilocal mass asymptotically behaves as
\begin{equation}
M \sim r.
\end{equation}         
We can relate the magnetic field $\hat{\cal B}$ in the orthnormal basis to $\cal B$ in the coordinate basis by ${\hat{\cal B}}=r^{-1}{e^{-\omega /2}}{\cal B}$. We define the magnetic charge by
\begin{equation}
Q_m=\lim_{r \to \infty}( r \hat{\cal B} )|_{a=0},
\end{equation}
where we switch off the effect of the dilaton field by setting $a=0$. We then obtain
\begin{equation}
Q_m={\sqrt{c_3}\over{2 \cosh\sqrt{c_3 \over 2}(z_2-z_3+ \sqrt{2 \over |c_2|}\pi)e^{{c_1 \over 2}(z_2-z_1+ \sqrt{2 \over |c_2|}\pi)}}}.
\end{equation}
The asymptotic behavior of dilaton field is 
\begin{equation}
e^{b\phi}\sim {1 \over r^2}.
\end{equation}
In the derivation of the present solutions we have assumed that there hold relations $2B=b^2$ and $4a/b=1$. 
They are satisfied by the string theory where $a=1$, $b=4$ and $B=8$.
The fact that we could solve the system under the relations might suggest the integrability of the string theory.

To conclude, we have exhausted and classified all the physical solutions to the 2+1 dimensional dilatonic Einstein equation with a magnetic charge for positive and negative cosmological constant under the static and rotationally symmetric assumption.
 One of the solutions possesses a horizon and has its validity outside the horizon. 
The magnetic field behaves asymtotically as $1/r$ and the dilaton field as $-{1 \above1pt 2}\log r$ when $b=4$.
 Besides the dilaton field, we also find that the mass diverges asymptotically as far as we suppress the contribution from the boundary.
We also find a solution with negative cosmological constant which does not have an event horizon, and so it is a particle-like solution. 
In addition to these solutions with negative cosmological constant, we obtain a solution with positive cosmological constant which is valid only inside of the horizon. 
The technique used here is also applicable to dilatonic Einstein equation with an electric field or both electric and magnetic fields. 
When we adopt the values of $a$, $b$ and $B$ obtained in the string theory, we can discuss the general solution without introducing any assumption besides the static and rotational symmetry assumption. This will be published in the forthcoming paper.

The authors would like to acknowledge the members of Doy\=o-kai for giving them a chance to start the present work.


\newpage


\begin{thebibliography}{99}

\bibitem{btz}
 M.Ba\~{n}ados, C.Teitelboim and J.Zanelli, Phys.Rev.Lett.{\bf 69} 1849 (1992) 
\bibitem{bhtz}
M.Ba\~{n}ados, M.Henneaux, C.Teitelboim and J.Zanelli, Phys.Rev.{\bf D48} 1506 (1993)

\bibitem{car}
S.Carlip, Class.Quantum Grav. {\bf 12} 2853 (1995)

\bibitem{howl}
G.T.Horowitz and D.L.Welch, Phys.Rev.Lett.{\bf 71} 328 (1993)

\bibitem{kal}
N.Kaloper,  Phys.Rev.{\bf D48} 2598 (1993)





\bibitem{kako}
M.Kamata and T.Koikawa,  Phys.Lett.{\bf 353B} 196 (1995); {\it ibid} {\bf 391B} 87 (1997)

\bibitem{cs}
M.Cataldo and P.Salgado,  Phys.Rev.{\bf D54} 2971 (1996), gr-qc/9605049

\bibitem{cm}
K.C.K.Chan and R.B.Mann, Phys.Rev.{\bf D50} 6385 (1994); erratum,{\bf D52} 2600 (1995) 

\bibitem{kp}
Y.Kiem and D.Park, Phys.Rev.{\bf D55} 6112 (1997)

\bibitem{hw}
E.W.Hirschmann and D.L.Welch, Phys.Rev.{\bf D53} 5579 (1996)

\bibitem{pk}
D.Park and J.K.Kim, J.Math.Phys.{\bf 38} 2616 (1997)


\bibitem{ky}
T.Koikawa and M.Yoshimura, Phys.Lett.{\bf 189B} 29 (1987)
 
\bibitem{by}
J.D.Brown and J.W.York, Phys.Rev.{\bf D47} 1407 (1993)

\bibitem{bcm}
J.D.Brown and J.Creighton and R.B.Mann, Phys.Rev.{\bf D50} 6394 (1994)


\end{thebibliography}
\end{document}